\documentclass[
superscriptaddress,
 amsmath,amssymb,
 aps,
twocolumn
]{revtex4-1}

\usepackage{graphicx}
\usepackage{dcolumn}
\usepackage{bm}

\usepackage{color} 

\begin{document}


\title{Long-range attraction of particles adhered to lipid vesicles}

\author{Raphael Sarfati}
\affiliation{
Department of Applied Physics
}
\affiliation{
Integrated Graduate Program in Physical and Engineering Biology, Yale University, New Haven CT 06511
}
\author{Eric R. Dufresne}%
 \email{ericd@ethz.ch}
\affiliation{
Department of Materials 
ETH Z\"{u}rich, 8092 Z\"{u}rich, Switzerland
}
\affiliation{
School of Engineering and Applied Sciences,
Yale University, New Haven CT 06511
}

\date{\today}

\begin{abstract}
Many biological systems fold thin sheets of lipid membrane into complex three-dimensional structures.
This microscopic origami is often mediated by the adsorption and self-assembly of proteins on a membrane.
As a model system to study adsorption-mediated interactions, we study the collective behavior of micrometric particles adhered to a lipid vesicle.
We estimate the colloidal interactions using a maximum likelihood analysis of particle trajectories. 
When the particles are highly wrapped by a tense membrane, we observe strong long-range attractions with a typical binding energy of 150~$k_B T$ and significant forces extending a few microns.

\end{abstract}

\pacs{Valid PACS appear here}
\maketitle


\section{Introduction}

The geometry of lipid membranes is essential to living cells.
Their topology defines the boundaries of the cell, nucleus, and organelles \cite{alberts}.
Their shape and size also play an essential role in cellular physiology, from the contraction of muscle \cite{ttubule} to the creation of vivid structural color \cite{butterfly}.   
Therefore, regulation of membrane geometry is of fundamental importance to cell biology.

In many cases, the folding of lipid membranes into complex three-dimensional structures is achieved by the adsorption and self-assembly of proteins on the surface of a lipid membrane \cite{bar}.
While many of the essential molecules have been identified, relatively little is known about the basic physics of protein-assisted membrane folding.  
Experiments have demonstrated a coupling between membrane curvature, tension, and binding affinity \cite{baumgart,hutchinson}.
Furthermore, membrane folding is intricately related to the organization of adsorbed proteins into supermolecular structures \cite{bar}.
These observations have inspired a number of theoretical studies considering the adsorption and interaction of proteins on membranes \cite{deserno2007,santangelo}.
However, experimental measurements of the interactions of membrane-bound proteins are unavailable.

The mechanics of bare lipid membranes is a  compromise of tension and bending energy \cite{canham,helfrich}. 
When particles adsorb, the physics is enriched by the particles' adhesion energy and geometry.
More precisely, for a piece of membrane of shape $\mathcal{S}$ bound to a particle, the energy of the system is described by the Helfrich Hamiltonian
\begin{equation}
\mathcal{H}(\mathcal{S}) = - w a_c + \int \left( \tau + \frac{1}{2}\kappa C^2 \right) dA ,
\end{equation}
where $\tau$ is the membrane tension, $\kappa$ the bending rigidity, $C$ the local total curvature, and $w$ and $a_c$ the adhesive surface energy and area of contact between the membrane and the particle.
An important material length scale emerges, $\lambda = \sqrt{\kappa/\tau}$.
Bending dominates on shorter scales, and tension dominates on longer ones.  
Bending rigidities of lipid bilayers are typically around 20~$k_B T$, so for moderately tensed vesicles ($\tau \sim 10^{-5} - 10^{-4}$~N/m), $\lambda \sim$ 50-100~nm \cite{tension}.

While membrane-mediated interactions of bound proteins are challenging to access experimentally, a few studies have made observations of membrane-induced attractions between micrometric colloidal particles \cite{angelova,safinya,kraft,stebe}.
These observations are not consistent with analytical theories of interactions of spherical particles which assume small deformations and predict repulsive interactions \cite{review}.  
On the other hand, numerical studies in the large deformation regime have found attractions between spheres \cite{deserno2007,deserno2011,cacciuto,mkrtchyan}.

Here, we investigate the interactions of membrane-bound particles using micron-sized colloidal particles attached to a giant unilamellar vesicle (GUV).  
When particles are highly wrapped by a tense membrane, they  spontaneously aggregate.
We describe a maximum likelihood analysis to estimate the pair potential from the approach to binding of individual particle pairs.
The  potential is strongly attractive ($>100~k_BT$ deep), and long-ranged ($>4~\mu \rm{m}$).

\section{Experimental results}

\begin{figure}[b]
\includegraphics[scale=0.33]{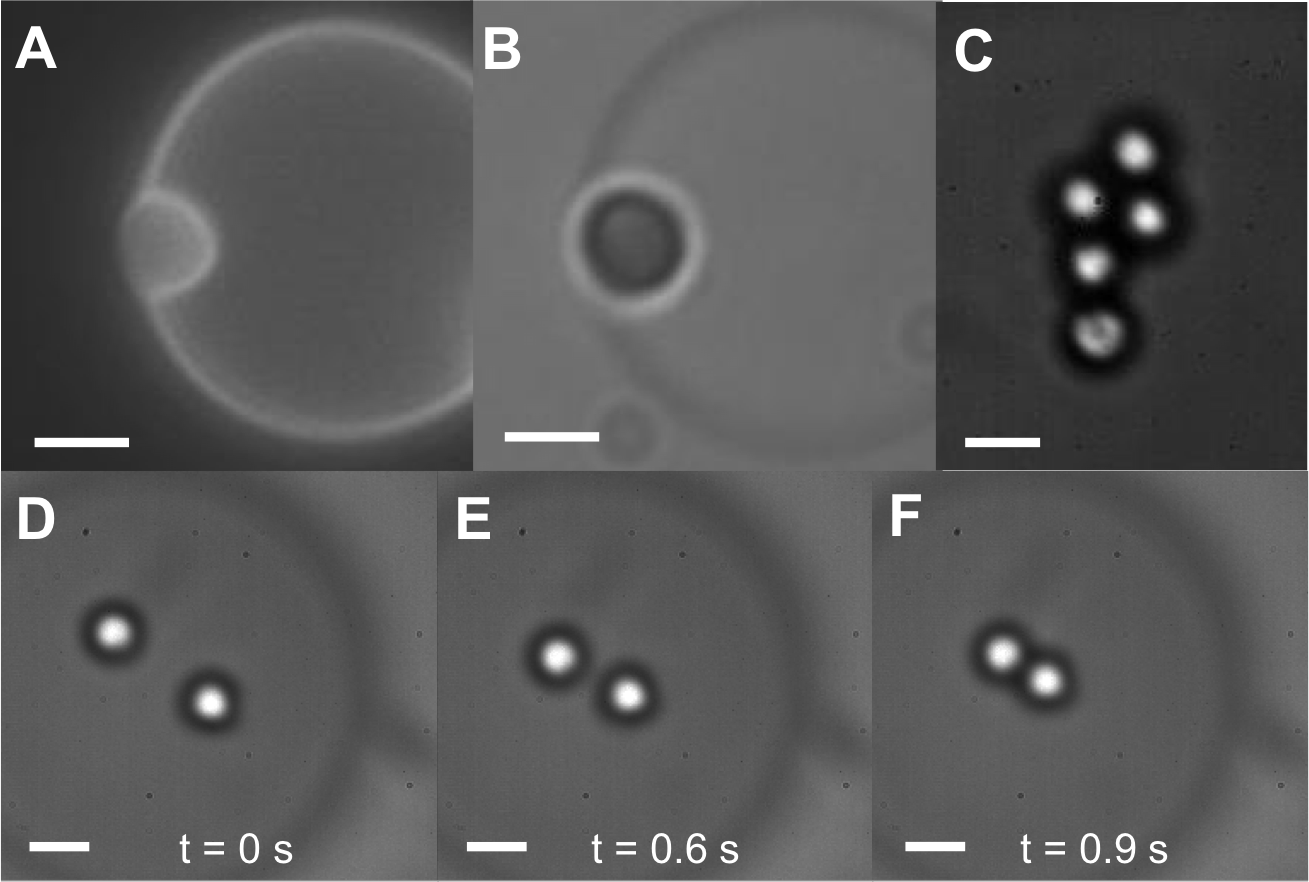}
\centering
\caption{\label{fig1} 
Particle binding and interaction on GUVs. 
Scale bar is 2~$\mu$m. 
(A) Fluorescent and (B) corresponding brightfield images of a bead strongly bound to the equatorial plane of a vesicle. 
More than half of the particle's surface appears to be wrapped. 
The GUV is tense, since its shape is spherical and no undulations are visible.
(C) Particles self-assemble when bound to the same GUV. 
(D-E-F) Time sequence of two particles at the top of a GUV (18~$\mu$m diameter) interacting across a distance of over 4~$\mu$m, and quickly moving towards each other in a time of about 1~s.
}
\end{figure}

Giant unilamellar vesicles of POPC (98\%), enhanced with lipids functionalized by rhodamine (1\%) and PEG-biotin (1\%), are fabricated by electroformation \cite{electroformation}.
They are re-suspended in a hypotonic buffer, and settle onto a non-adherent coverslip.
The vesicles have a wide range of tensions: 
some exhibit large shape fluctuations, while others are smooth and nearly spherical (Fig. \ref{fig1}A).
Using optical tweezers (1064 nm), we bring streptavidin-functionalized polystyrene spheres (radius $R=1~\mu$m) in contact with GUVs of diameter from 15~$\mu$m to 20~$\mu$m.
There is strong adhesion of the particles to the bilayer due to the interaction of biotin with streptavidin.
The extent of adhesion varies somewhat from bead to bead, but the membrane typically wraps the bead past its equator, as shown in Fig. \ref{fig1}A.

Even though individual particles are stable in the bulk, beads bound to tense GUVs formed clusters, as shown by the micrograph in Fig. \ref{fig1}C.
Thermal fluctuations were not able to dismantle these clusters, but they did cause significant fluctuations in the particle separation, as we will discuss later.
Beads bound to flaccid GUVs did not aggregate.

\begin{figure}[b]
\centering
\includegraphics[width=\linewidth]{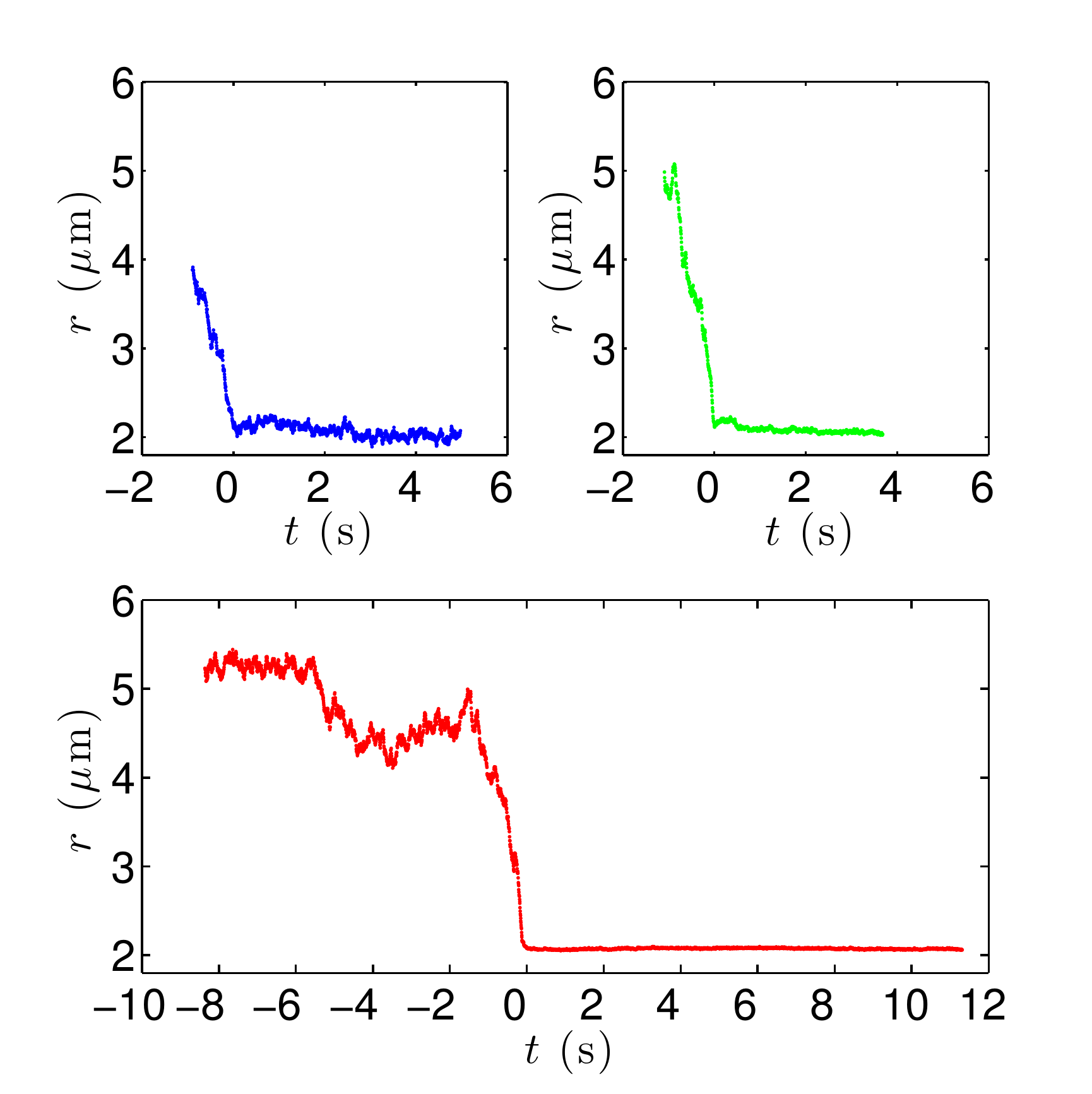}
\caption{\label{fig2} 
Representative trajectories of particle pair separations on three different vesicles.
All trajectories equilibrate at a distance around 2~$\mu$m in the bound state ($t \geq 0$~s). 
}
\end{figure}

These particle interactions are long-ranged.
Particles within a few microns of one another move quasi-ballistically toward a bound state, as shown in the time sequence in Fig. \ref{fig1}(D-E-F).
We imaged the approach of particle pairs with a high-speed camera (250 frames per second) and extracted the bead positions using a standard particle tracking algorithm \cite{crocker}.
Three representative trajectories of the center-to-center separation between two 2-$\mu$m-diameter beads bound to the same vesicles are shown in Fig. \ref{fig2}. 
All show a strong attraction starting from over 4~$\mu$m away, with average velocities around 1~$\mu$m/s, and significant fluctuations about a bound state near contact.
A movie showing the particles motion is available in the Supplemental Material.

\section{Maximum likelihood analysis}

We aim to quantify these membrane-mediated interactions.
Analysis of the fluctuations near equilibrium \cite{frej} enables measurement of the stiffness of the particle-particle bond, but they do not sample the long-range interaction.
In principle, the long-range interaction could be probed using optical tweezers as a force transducer \cite{transducer}, or blinking optical tweezers \cite{crocker,sainis}.
Unfortunately, we have found that lipid vesicles are perturbed by the laser traps \cite{angelova,moroz}; however, they relax after about 200~ms after the laser is blocked.

Consequently, we introduce an alternate approach to quantify interaction parameters from individual trajectories, based on the general method of maximum likelihood \cite{ml}.
Consider a Brownian particle moving in one dimension, with position $x$.
Its dynamics are given by the Smoluchowski Equation \cite{sainis}.
In general, the diffusion coefficient $D$, and  applied force $F$, may depend on~$x$.
However, over sufficiently short time intervals $\Delta t$, the particle samples a region where force and diffusion coefficient are uniform.
In this case, the change in the particle position, $\delta~=~x(t+\Delta t)-x(t)$, is given by a Gaussian probability distribution $p(\delta \, \vert \, \Delta t, F(x), D(x))$. 
The mean, $\mu$, and standard deviation, $\sigma$, of the distribution depend on the force and diffusion coefficient as
\begin{equation}
\mu = \left( \frac{FD}{k_B T} + \frac{dD}{dx} \right)\Delta t,
\label{mu}
\end{equation}
\begin{equation}
\sigma = \sqrt{2D\Delta t}.
\label{sigma}
\end{equation}

Consider a discretely sampled one-dimensional trajectory $\{x_1, ..., x_N\}$, where $x_i$ indicates the  coordinate at time $t_i$. 
The force and diffusion profiles are unknown, but we assume that they can be by described by a discrete set of parameters $\alpha^0 = \{ \alpha^0_1,...,\alpha^0_q \}$. For example, in the case of homogeneous force and mobility, this is simply $\{F_0,D_0\}$ where $F_0$ and $D_0$ are constants. 
Given a trial set $\alpha = \{ \alpha_1,...,\alpha_q \}$, the probability density of observing the trajectory $\{x_1, ..., x_N\}$ is
\begin{equation}
\mathcal{P}( \alpha )=\prod_{i=1}^{N-1} p(\delta_i \, \vert \, x_i, \Delta t, \alpha ),
\label{prob}
\end{equation}
where $N$ is the number of sampled timepoints, and the discrete displacements are $\delta_i=x_{i+1}-x_{i}$.
In the limit of large trajectories ($N \rightarrow \infty$), $\mathcal{P}( \alpha )$ is maximum when $\alpha = \alpha^0$
\cite{sarfati}.
In practice, the analysis is more stable numerically if one maximizes a log-likelihood function
\begin{equation}
\mathcal{L}( \alpha ) = \frac{1}{N-1} \sum_{i=1}^{N-1} \ln \left[ \, p \left( \delta_i \, \vert \, x_i, \Delta t, \alpha  \right) \, \right]
\end{equation}

The main benefit of this approach is that it does not require the construction of an empirical probability distribution \cite{merrill}, and can be implemented with a single trajectory.
The key limitation of this approach is that it requires a model for the spatial dependence of the force and diffusion coefficient.

\begin{figure}[b]
\includegraphics[width=\linewidth]{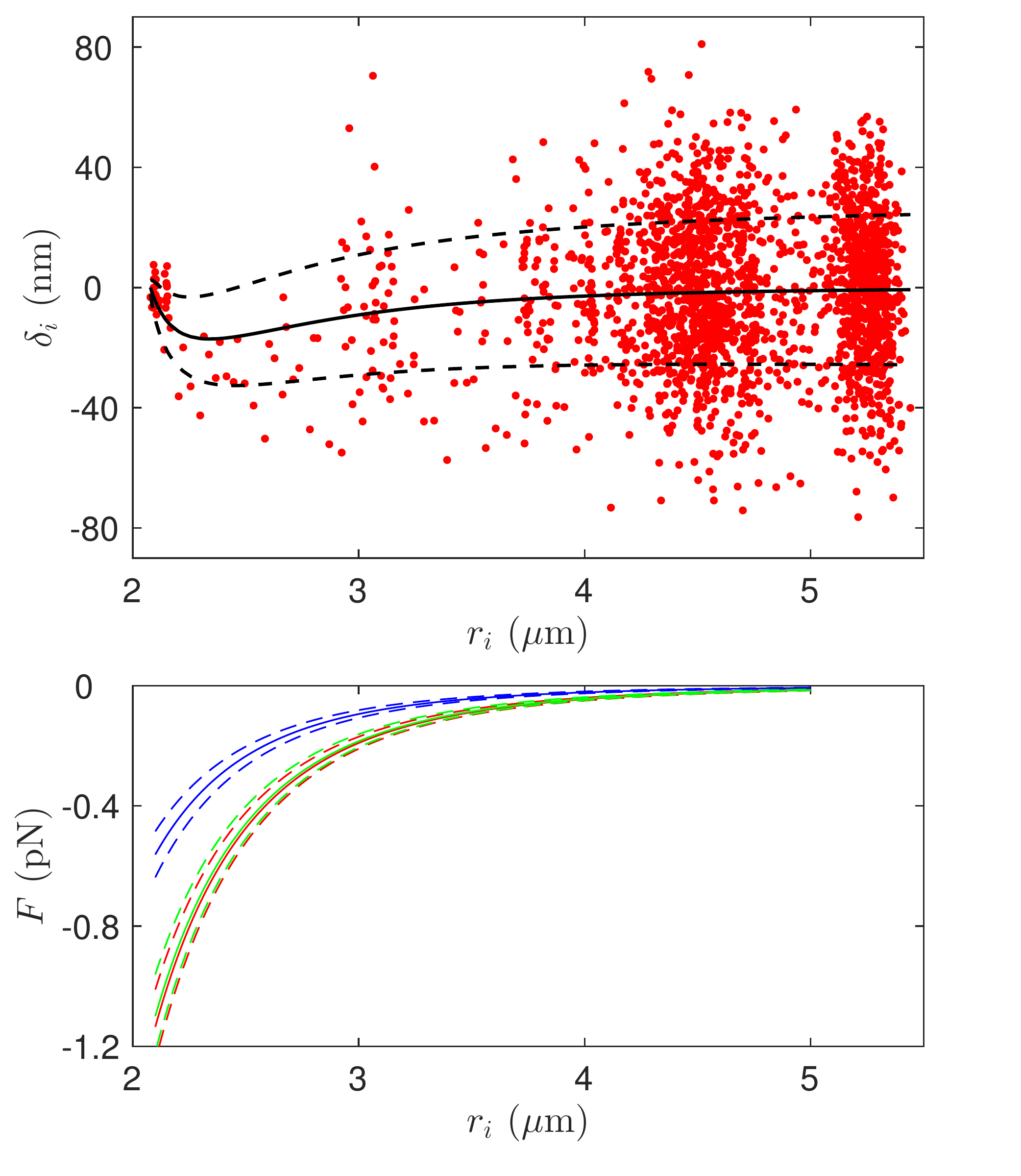}
\caption{\label{fig3} 
Displacement and force fits.
(Top) Frame-to-frame ($\Delta t$~=~4~ms) displacements $\delta_i$ as a function of pair separation $r_i$ for the far-field part of the red trajectory. The red dots represent the experimental data. The solid black line is the maximum likelihood fit for the mean displacement. The dashed lines represent the typical fluctuations due to Brownian motion ($ \sqrt{2D(r)\Delta t}$).
(Bottom) Force profiles obtained from the maximum likelihood analysis of the blue, green, and red far-field trajectories. The dashed lines represent the statistical uncertainty on the fits (see Supplemental Material Text and Fig.~S5).
}
\end{figure}

We apply the maximum likelihood analysis to estimate the interaction parameters for the three trajectories shown in Fig. \ref{fig2}. 
We start by restricting our attention to the far-field attractive interaction, laying aside the stably bound portion of the trajectory.
As an example, the displacements as a function of interparticle separation for the red trajectory are presented in Fig. \ref{fig3}(Top). 

It is important to note that, to our knowledge, there is no definite theoretical form for the attractive force between micron-sized particles on a tense vesicle \cite{review}. 
A new hypothesis, recently introduced in \cite{stebe}, is that the contact lines between the membrane and the beads are pinned \cite{dietrich} in a complex geometry.
The basic idea is that rough contact lines deform the membrane  and induce attractive interactions, analagous to those between particles at the interface of two simple fluids.  
These interactions have been formalized in \cite{stamou} and the dominant term is quadrupolar, so that the force is predicted to be a power law with exponent $-5$.
This is the description which we adopt here, so we assume that the force between the particles has the form
\begin{equation}
F(r) = -\phi \frac{1}{r^5}.
\end{equation}
Therefore, the force $F$ depends on one parameter, $\phi$.

Similarly, there is no appropriate theory for the relative diffusion coefficient of two large beads bound to a lipid membrane, where both the liquid and the membrane contribute \cite{ruocco,schwille,pincus,parthasarathy}.
Therefore, we assume a simple form for the relative diffusion coefficient:  
it should be zero in contact and plateau to some constant value at large separations.
These basic criteria are satisfied by the form for identical spheres in a viscous fluid \cite{biancaniello}:
\begin{equation}
\label{diffusion}
D(r) = D_0 \times \dfrac{12(r/R_0-2)^2 + 8(r/R_0-2)}{6(r/R_0-2)^2 + 13(r/R_0-2) + 2},
\end{equation}
where $D_0$ is the 
one-particle
diffusion coefficient at infinite separation, and $R_0$ is the hydrodynamic radius of the particle.
In a viscous fluid, $D_0$ and $R_0$ are related through the original Stokes-Einstein relation.
Here, we let them vary independently to accommodate contributions from both the membrane and bulk.
Putting together these forms for the force and hydrodynamic drag, 
the far-field trajectories are characterized by a set of three parameters: $\{ \phi, D_0, R_0 \}$.

We report the parameter values that maximize the likelihoods for the trajectories in Fig. \ref{fig2} in Table \ref{ff_results}.
The inferred mean and standard deviation of the frame-to-frame displacement distribution for the red trajectory are plotted on top of the datapoints in Fig. \ref{fig3}(Top) and for the others in the Supplement.
The inferred force profiles for all three trajectories are shown in Fig. \ref{fig3}(Bottom).
The forces have maximum values of about 1~pN, and decay over lengthscales of a few microns.

\begin{table}[b]
\caption{\label{ff_results}
Maximum likelihood estimates of the trajectories force and diffusion parameters in the far-field and near-field. 
The number of significant digits comes from numerical uncertainty in maximization of the likelihood.
}
\renewcommand{\arraystretch}{1.2}
\begin{tabular}{l|c|c|c|}
\textrm{}&
Blue &
Green &
Red \\
\hline
$\phi$ ($\times 10^{-41}$) & 2.29 & 4.49 & 4.64  \\
$D_0$ ($\times 10^{-14}$ m$^2$/s) & 8.63 & 9.88 & 5.48 \\
$R_0$ ($\mu$m)& 1.046 & 1.050 & 1.038 \\
\hline
$k$ (nN/$\mu$m) & 0.79 & 9.4 & 79.5 \\
$r_{eq}$ ($\mu$m) & 2.06 & 2.071 & 2.074 \\
\hline
\end{tabular}
\end{table}

The force profiles for the three pairs of particles appear quite different.
To determine if these differences are significant, we investigated the robustness of these force profiles, and their associated uncertainties.
We quantified the statistical uncertainty in a force profile using numerical simulations of Brownian trajectories, and extracted the 25th and 75th percentiles from the distributions of their maximum likelihood force fits.
We report the corresponding confidence intervals as dashed lines in Fig. \ref{fig3}, which correspond to roughly $\pm15\%$ of the input value (see Appendix).
We tested for sensitivity to the assumed function form by analyzing the data with various functional forms of the force.
The recovered force profiles are very similar to the ones presented in Fig. \ref{fig3}.
Details are provided in the Supplement.
Therefore, the differences in the force profiles between the blue trajectory and the green and red appear as significant, and presumably due to differences in the membrane tension and the wrapping of the bead by the membrane.

The spatial dependence of the diffusion coefficient is surprisingly well-captured by  Eq. \ref{diffusion}, with a hydrodynamic radius, $R_0$, that is not significantly different from the particle radius, $R$, in any of the three trajectories (Table \ref{ff_results}).
However, the limiting value of the diffusion coefficient varies more significantly from particle to particle, perhaps reflecting differences in the extent of wrapping by the membrane.

\section{Near-field interaction}

Having analyzed the far-field attraction, we now focus on the interaction in the bound state.
Significant separation fluctuations are observed, with essentially two characteristic timescales.
Notably in the first 0.5~s of the bound state, we notice some slow features (see Fig. \ref{fig2}, blue and green), which we attribute to evolving wrapping of the membrane around the particles. 
After this transition period, the particle separation fluctuates about an apparent equilibrium separation, with a characteristic time of about 5~ms.
At this point, the particle interaction should be given simply by Hooke's law, $F(r) = -k(r-r_{eq})$, with $k$ the spring constant and $r_{eq}$ the equilibrium distance.

We apply the maximum likelihood analysis to estimate these parameters for each trajectory, and report our results in Table \ref{ff_results}.
As expected, the equilibrium separation of the particles is consistently found to be close to the nominal particle diameter.  
Interestingly, the measured spring constants span two orders of magnitude.
We suspect that these widely varying stiffnesses depend strongly on the tension and state of wrapping  of the particles by the membrane.
In Appendix B, we show that these estimations are very similar to estimations of $k$ and $r_{eq}$ obtained from standard Boltzmann statistics analysis of these trajectories.

\begin{figure}[b]
\includegraphics[width=0.8\linewidth]{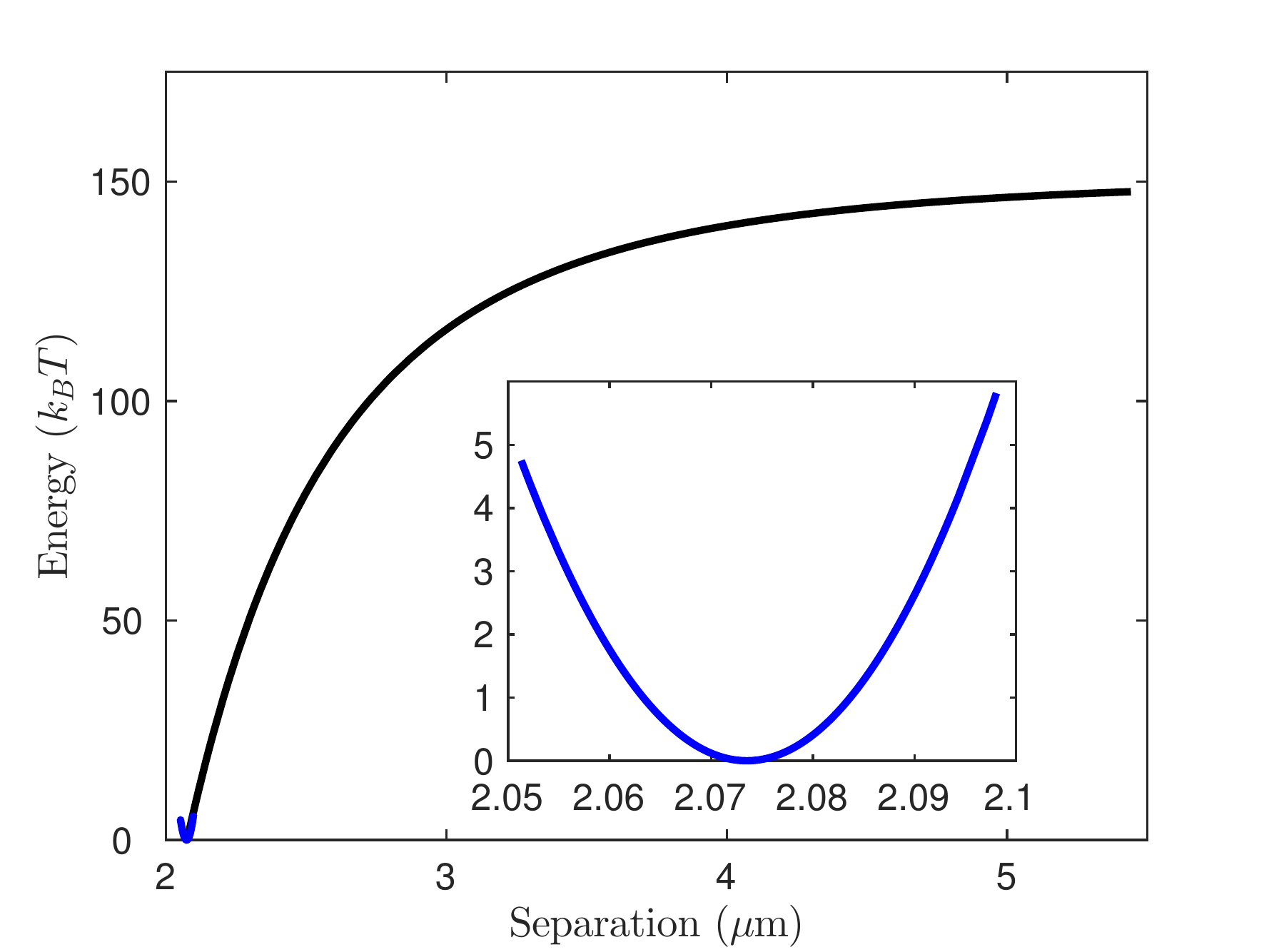}
\caption{\label{fig4}
Plot of the estimated energy landscape as a function of the pair separation. Inset: zoom-in of the near-field part. The binding energy is about 150~$k_B T$.
}
\end{figure}

By integration of the measured force profile, we construct the two-particle membrane-mediated energy landscape for the red trajectory in Fig.~\ref{fig4}.
The potential depth approaches 150~$k_B T$.

\section{Conclusion}

We observed long-range attractions between micron-sized functionalized polystyrene spheres strongly adhered to a tense lipid bilayer.
We estimated pair interactions based on a maximum likelihood analysis.
This approach estimates the force profile with many fewer observations than spatially resolved measurements of the drift velocity and diffusion coefficient \cite{sainis}, and therefore is well-suited to single-trajectory analysis.

While there are many possible origins for the observed long-range attraction, tension mediated interactions seem to be the most likely candidate.
Tension-mediated interactions are analogous to capillary interactions of particles at a fluid interface.
Since the weight of the particles is too small to induce significant membrane deformations \cite{kralchevsky,nicolson}, the most plausible origin of the long-range attraction is  multipolar capillary interactions due to roughness of the contact line. 
Previous studies have mentionned the possibility of a pinned contact line between a sphere and a lipid membrane \cite{dietrich}, which has recently been suggested to be the origin of attraction of particles on lipid membranes \cite{stebe}.
Multipolar capillary interactions are well understood at fluid-fluid interfaces, and have a dominant term corresponding to the quadrupolar mode \cite{stamou},  Eq. 6.
In order to elucidate the exact mechanism behind these long-range interactions, further measurements quantifying and controlling the tension in the membrane are necessary.


\section*{Acknowledgments}
The authors thank Jason Merrill, Jin Nam, Frederic Pincet, Rob Style, and anonymous reviewers for helpful suggestions. This work was supported by the National Science foundation (CBET 12-36086).

\section*{Appendix A: Materials and Methods}
Phospholipids were purchased from Avanti Lipids: 
1-palmitoyl-2-oleoyl-sn-glycero-3-phosphocholine (POPC), 
\linebreak
1,2-disteoroyl-sn-glycero-3-phosphoethanolamine-N- [biotinyl (polyethylene glycol) 2000] (DSPE-biotin), and 
\linebreak
L-$\alpha$-phosphatidylethanolamine-N-(lissamine rhodamine B sulfonyl) (PE-Rhod).
Lipid vesicles where synthesized using the electroformation method \cite{electroformation}. 
Briefly, 50~$\mu$L of a mixture of POPC/DSPE-biotin/PE-Rhod 98:1:1 (1~mg/mL in chloroform) were deposited using a glass syringe (Hamilton Gastight) onto two platinum wires contained in a teflon chamber.
The chamber was filled with 1.6~mL of a solution of 200~mM of sucrose, then sealed, and the wires where connected to a signal generator (Wavetek FG2~A) applying a sinusoidal voltage (10~Hz, 8~V peak-to-peak) for 2-to-8~hours.
A working solution was made by mixing 10~$\mu$L of the lipid vesicle solution, 3~$\mu$L of streptavidin coated, 2~$\mu$m diameter latex beads (Polysciences, volume fraction 1.36\%), and 87~$\mu$L of a hypotonic binding buffer solution (62.5~mM KCl, 25~mM glucose, 12.5~mM HEPES, 0.5\% Bovine Serum Albumine).
After gentle homogenization, 7~$\mu$L of the solution were deposited in a sealed, thin chamber (Secure-Seal spacer from Life technologies, 9~mm diameter and 120~$\mu$m thickness) in order to prevent any flow by turbulence or evaporation.
Observation was realized using an inverted microscope (Nikon TE-2000), equipped with a fluorescent filter and a N.A. 1.4 100x oil immersion objective lens.
Movies were recorded using a fast camera (Photron Fastcam 1024PCI).
The beads were manipulated using a holographic optical tweezers setup described in \cite{trapsetup}.

The particle positions were extracted from the movies using a standard particle tracking algorithm in \textsc{Matlab} \cite{crocker}.
The maximum likelihood analysis was performed using the \textit{fminsearch} function in \textsc{Matlab}, which finds the position of the minimum of a scalar function of several variables. This function requires to input an initial guess for the position of the minimum. We made sure that our results were independent of the guess inputs by trying several dozens of initial guesses over a wide range of parameters.

\section*{Appendix B: Numerical investigation of maximum likelihood analysis}

\subsection{Far-field simulations}

We perform numerical simulations to investigate the robustness of the maximum likelihood analysis that we use in this article, and estimate the associated uncertainties.

We simulate the motion of a Brownian particle with diffusion coefficient $D(x)$ in a force field $F(x)$, both dependent on the particle's position $x$. 
We analyze these trajectories using the maximum likelihood analysis.

To mimic our experimental trajectories, we generate $N_s = 1000$ random trajectories of 300 points separated by a time interval $\Delta t$ = 4ms and all finishing at $x$ = 2.1~$\mu$m . 
These trajectories were obtained from longer trajectories simulated at a higher frequency of $10^4$ frames per second. 
We use the force profile
\begin{equation}
F(x) = -\phi \frac{1}{x^5},
\end{equation}
and the diffusion profile
\begin{equation}
D(x) = D_0 \times \dfrac{12(x/R_0-2)^2 + 8(x/R_0-2)}{6(x/R_0-2)^2 + 13(x/R_0-2) + 2},
\label{diffusion_profile}
\end{equation}
with parameters values 
for $\phi$, $D_0$, and $r_0$ corresponding to the values extracted from our trajectories fits.
We apply the maximum likelihood analysis, and compare the results to these values.

\begin{figure*}[t]
\includegraphics[width = \linewidth]{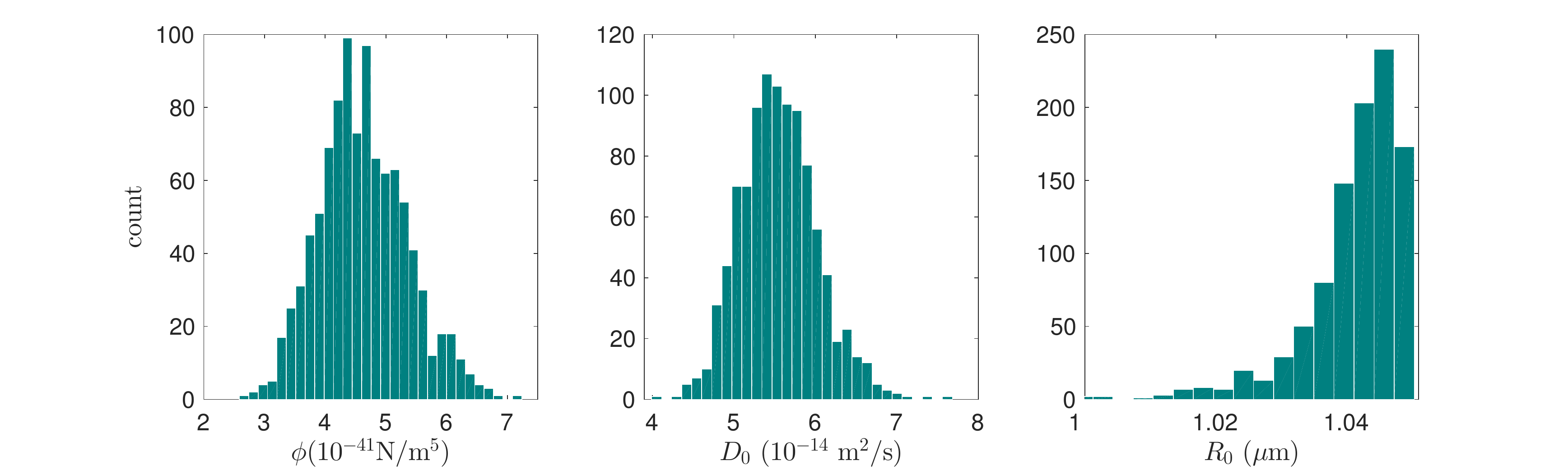}
\caption{\label{hist3} 
Histogram of the results of the maximum likelihood analysis for $\phi, D_0, R_0$ for each of the 1000 simulated trajectories.
}
\end{figure*}

\begin{figure*}[t]
\centerline{
\includegraphics[scale = 0.5]{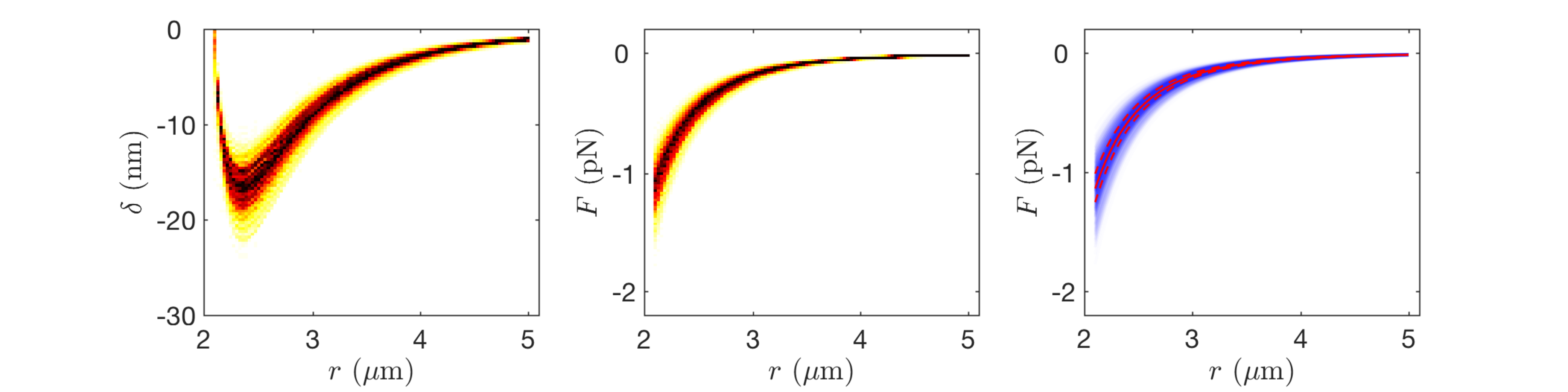}
}
\caption{\label{displacementsforcefits} 
Maximum likelihood fit distributions.
(Left) Displacement fits. 
The high density area (darker colors) correlates to the input displacement profile.
(Middle) Force fits.
Again, the high density area correlates with the input force profile.
(Right) Force fits for all simulated trajectories (blue curves), and input force profile (solid red curve).
The 25th and 75th percentiles (dashed red curves) are obtained from the distribution shown in the middle plot.
}
\end{figure*}

The parameter values estimated from the maximum likelihood analysis are presented as histograms in Fig.~\ref{hist3}, for the simulations corresponding to the red trajectory ($\phi = 4.64 \times 10^{-41}$~N/m$^5$, $D_0 = 5.5 \times 10^{-14}$~m$^2$/s, $r_0 = 1.04 \times 10^{-6}$~m).
We find that estimations of the three parameters are closely distributed around their actual values.
As a consequence, the force and displacement profiles are very similar, as seen in Fig.~\ref{displacementsforcefits}.

These simulated trajectories allow us to estimate the uncertainty on our estimated parameters. 
From the $N_s$ estimated force profiles, we can calculate the $25^{th}$ and $75^{th}$ percentiles from the force distribution at each separation in order to estimate the 50\% confidence interval, which we reported in Fig.~3, as shown in Fig.~\ref{displacementsforcefits}.

\subsection{Near-field comparison between Boltzmann statistics and maximum likelihood}

In the near-field, where the particle separation fluctuations about an equilibrium position, the spring constant $k$ and equilibrium distance $r_{eq}$ can be infered both from the maximum likelihood method and from a more traditional Boltzmann statistics analysis.
Here, we present a comparison of the two methods.

Following Boltzmann statistics, we can calculate the spring constant $k^{BS}$ from 
a set of experimental separations $\mathcal{X} = \lbrace x_1, ..., x_{n} \rbrace$ 
using:
\begin{equation}
k^{BS} = \frac{k_B T}{\sigma^2(\mathcal{X})},
\end{equation}
where $\sigma^2$ denotes the variance.


In Table \ref{bs_ml}, we present a comparison of the results obtained from Boltzmann statistics and from the maximum likelihood analysis. 
The results agree very well.

\begin{table}[h]
\caption{\label{bs_ml}
Results from the maximum likelihood (ML) analysis, and comparison with the Boltzmann statistics analysis (BS). 
The number of significant digits comes from numerical uncertainty in maximization for ML, and standard statistical uncertainty for mean and variance for BS.
}
\begin{tabular}{l|c|c|c|}
\textrm{}&
Blue &
Green &
Red \\
\hline
$k^{ML}$ (nN/$\mu$m) & 0.79 & 9.4 & 79.5 \\
$k^{BS}$ (nN/$\mu$m) & 0.76 & 8.99 & 72.3 \\
\hline
$r_{eq}^{ML}$ ($\mu$m) & 2.06 & 2.071 & 2.074 \\
$r_{eq}^{BS}$ ($\mu$m) & 2.06 & 2.07 & 2.07 \\
\hline
\end{tabular}
\end{table}

%

\end{document}